# SMART GRIDS SECURED BY DYNAMIC WATERMARKING: HOW SECURE?

KATE DAVIS [1], LASZLO B. KISH [1,2,†], CHANAN SINGH [1]

[1]*Department of Electrical and Computer Engineering, Texas A&M University, TAMUS 3128, College Station, TX, USA*

[2]*Óbuda University[‡], Budapest, Bécsi út 96/B, Budapest, H-1034, Hungary*

Unconditional security for smart grids is defined. Cryptanalyses of the watermarked security of smart grids indicate that watermarking cannot guarantee unconditional security unless the communication within the grid system is unconditionally secure. The successful attack against the dynamically watermarked smart grid remains valid even with the presence of internal noise from the grid. An open question arises: if unconditionally authenticated secure communications within the grid, together with tamper resistance of the critical elements, are satisfactory conditions to provide unconditional security for the grid operation.

## 1. Introduction

Smart grid security [1-3] is important to maintain the confidentiality, integrity, and availability of smart grid infrastructures by preventing threats that can lead to business and operational disruptions. It involves safeguarding resources from unauthorized access, modifications, and ensuring the availability of the grid. Cybersecurity in the smart grid is essential due to the increased use of IT-based electric power systems, which increases cybersecurity vulnerabilities, leading to the need for resilient cybersecurity solutions. It is also important to protect smart grids against potential terrorist threats, espionage threats, and vulnerabilities caused by natural disasters, equipment failures, and user errors. The security of the smart grid involves ensuring that all risks are managed when things go wrong, making it a learning process and essential for the overall resilience of the smart grid. When supplier-managed demand response is employed, customer privacy and cybersecurity become crucial concerns for both the customer and the supplier. This is also true for the communication of meters to accountants.

In 2016, a novel technique known as Dynamic Watermarking (DW) was introduced [4] to protect the integrity of control systems within networked cyber-physical systems. Dynamic watermarking [4] in the context of smart grids involves the addition of continuous noise to the control system of the grid to enhance its security. This technique is used to detect and defend against cyber-physical attacks in the smart grid [4-6]. By adding a dynamic watermark (noise) to the control signals and by analyzing the response, any unauthorized modifications or attacks can be detected, ensuring the integrity and security of the grid's operations.

Nonetheless, a detailed cryptanalysis of DW has exposed, see in Section 4, that its security relies on the premise of a "restricted adversary", with considerably constrained capability. Consequently, the security provided by DW is conditional, indicating that it is vulnerable to being compromised and does not guarantee future-proof security.

To demonstrate this vulnerability, we detail a malicious attack that is effective against systems employing DW.

---

[†] Corresponding author
[‡] Honorary faculty.

First for the benefit of the Reader, we summarize some of the elements of security [7] in communication systems that are necessary for this analysis.

*1.1 Information-theoretic versus unconditional security in communications*

Table 1 summarizes the features and differences regarding conditional and unconditional security, respectively.

1.1.1 Conditional security

Nowadays, shared keys are typically obtained by number theoretical protocols that involve two communicators (Alice and Bob) exchanging data. These conditionally secure schemes are *not future proof*, because an eavesdropper (Eve) can crack the key with enough time and/or computational power. From an information-theoretic perspective, the key has zero information entropy for Eve, which means *zero security*. The security of these protocols relies on the *assumption* that it is computationally hard (but not impossible) to extract the key from the data exchange between Alice and Bob. However, this assumption has no mathematical proof; it is only based on the common intuitive opinions of many mathematicians. Effective algorithms utilizing polynomial computing power to crack the secure scheme may still exist and someone may find them. Consequently, these protocols are termed conditionally secure, contingent upon the validity of the assumption that breaking them indeed requires exponential computational power. Moreover, some of these protocols could be broken by a quantum computer, if such a device becomes available.

Table 1.

|  | **Conditional security** | **Unconditional security** |
|---|---|---|
| **Key features** | • Limited Eve/Trudy. Some examples: limited computational power; limited sensitivity or resolution; limited sampling rate; limited number of probing locations, etc.<br>Example: digital communications secured by math complexity based protocols:<br>• Security claim is based on an unproven assumptions, e.g.:<br>• The hardness of the underlying math problem is exponential;<br>• Eve is limited to using polynomial computing power, which excludes the use of quantum computing, noise-based logic, and other exponential algorithms;<br>• Not Future Proof: The information is susceptible to being accessed, and it's only a matter of time or technological advancement. | • Unlimited Eve/Trudy; Some examples: unlimited computational power; unlimited sensitivity or resolution; unlimited sampling rate; unlimited number of probing locations, etc.<br>• They are limited only by the laws of physics and certain elements of the protocol set by Alice and Bob (e.g. the active time period of a secure bit exchange);<br>• Future Proof: the information cannot be accessed by Eve; future technology improvements do not help to access communications in the past present of future. |
| **Comments** | • Tamper resistance for Alice's and Bob's private spaces are required.<br>• Cheap, software based. | • Tamper resistance for Alice's and Bob's private spaces are required.<br>• More expensive, hardware based. |

1.1.2 Unconditional security

In contrast, *information-theoretic security* or *unconditional security* [7-9] means that the key has maximum entropy for Eve, regardless of her computational power or physical limitations. During an arbitrary attack, for a key length of *N* bits, Eve's information entropy remains *N* bits, which means the key is secure even if Eve has unlimited resources that are restricted by only the laws of physics. Time and cost are irrelevant.

Ciphers that are utilizing the secret keys can be made unconditionally secure without hardware components, for example, Shannon's One Time Pad (Vernon cypher) [11]. However, secure key exchange is different. Currently there are only two unconditionally secure key exchanger families:

- The KLJN scheme [10,11] that is based on the *Second law of thermodynamics*.

- Quantum Key Distribution (QKD) [12,13], which is based on the *No-cloning theorem* of quantum informatics and/or *quantum entanglement*.

Both the KLJN and QKD schemes are complex and costly to implement. Nonetheless, they remain the only known methods for achieving unconditional security in communications.

*1.2 Authenticated communications*

Authenticated communication is used to verify the identity of the communicating parties. The communication is *not encrypted* and is available for the public. It is needed for both the QKD and KLJN schemes, in the public channel between Alice and Bob, for basic function (QKD) and/or to secure the systems against active attacks (QKD, KLJN). For an $N$-bit long message a small, $O[\log (N)]$, part of the secure key is used up for the communication of the digital signatures that must be encrypted. If the secure key fragment utilized for authentication is obtained by an information-theoretic (unconditional) key exchange then we call such an authentication unconditional.

*1.3 Kerckhoffs's principle (Shannon's maxim) of security*

To state that a communication system is unconditionally secure, it must satisfy the Kerckhoffs's principle (Shannon's maxim) [7], which means the adversaries know the system and the detailed protocol, except the secure key. The blueprint of the system and other protocol details are assumed to be known by adversaries. Therefore, keys must be spontaneously generated, used and annihilated.

*1.4. Digital twins*

Digital twins [14] are not strictly security matter however they are closely related to grid security because a digital twin can be constructed due to the Kerckhoff's principle and the lack of unconditionally secure communications in an unsecured system. It is a virtual model of a physical object or system, updated in real-time. Digital twins are more than replicas; they're evolving databases that mirror and inform their physical counterparts.

Note, a digital twin has also reference inputs [14]. As a digital twin is a digital model of a real-world physical system that is updated with inputs of real-world data from its physical counterpart.

**2. Classifying smart grid security: Conditional and unconditional security**

Here we classify smart grid security into the two fundamental classes of conditional and unconditional security, similarly to communications. The eavesdropper Eve now becomes troublemaker Trudy. Though, smart grids utilize communications, they are much more than just communicators, so there are differences. Table 2 generalizes the fundamental security classes of communications to smart grid security by extracting the essential properties of communication security and projecting them to the case of smart grids.

Table 2.

| | Conditional security | Unconditional security |
|---|---|---|
| **Key features** | **Assumed:**<br>• Limited Trudy;<br>• Security claim is based on Trudy's limitations, e.g.:<br>• Trudy does not have enough resources or cannot utilize them due to economical or other limitations;<br>• And/or Trudy does not have an accurate model of the system;<br>• And/or Trudy cannot exploit all the means that are available for her. | **Assumed:**<br>• Unlimited Trudy: she is limited only by the laws of physics;<br>• The security is maintained even if Kerckhoffs's principle holds: Trudy knows all the deterministic components, their structure and transfer functions in the system, and the details of the protocols.<br>• Trudy can exploit all the means that are available for her to attempt cracking the security. |
| **Comments** | • Tamper resistance for Alice's and Bob's private spaces are required. | • Tamper resistance for Alice's and Bob's private spaces are required.<br>• Unconditionally-authenticated communications with unconditionally secure key exchange between controls sensors and actuators are required, see the present study. |

Concerning the Comments row of Table 2, we point to the rest of this paper, particularly to Section 4.

**3. On the enhanced smart grid security offered by the watermarking method**

The idea of watermarking based security [4] is very simple. The Controller (Conrad) of the smart grid injects a small private random noise into the feedback loop of the control system. The smart grid transfer function is known by Conrad thus he can calculate the expected changes in the remote sensor signals. If this signal change differs from the expected one at one of more sensors, that is an indication that the integrity of the smart grid is compromised.

With focusing on the control and sensor units, a simplified smart grid illustration is shown in Figure 1. The feedback loops are closed by the lines from the sensors to the control inputs.

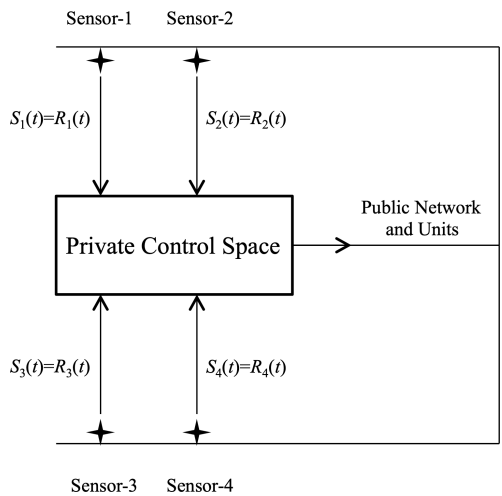

Figure 1. Illustration of the core of a smart grid control loop. In the ideal case, the sensor signal $S_i(t)$ is equal to its regular value $R_i(t)$, see below.

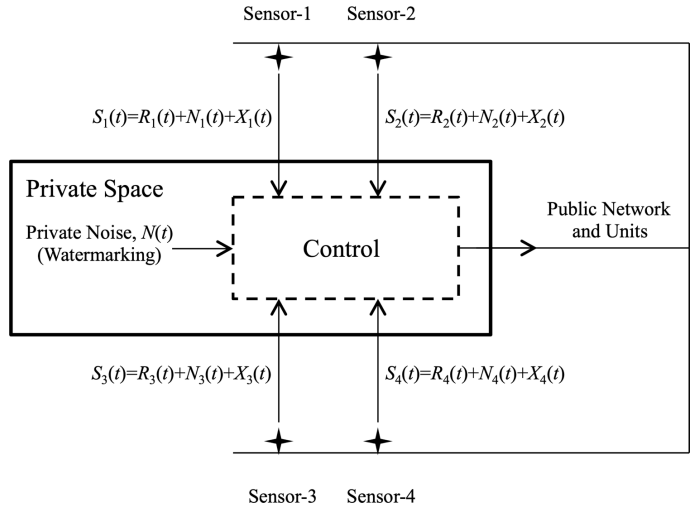

Figure 2. Watermarking and its impact on the sensor signals. In the linear situation, the sensor signal $S_i(t)$ is the sum of the regular sensor signal $R_i(t)$ and the $N_i(t)$ which is the watermarking signal after linear operations. When nonlinearity cannot be neglected, the nonlinear component $X_i(t)$, including cross terms with the original sensor signal also appear. Note: in practical situations, nonlinearities should be negligible to avoid undesired interactions between the watermarking signal and the normal operation of the grid.

Conrad can use various measures to detect attacks by a limited Trudy. For example, the variances of the watermarking related signal component $N_i(t)$ ; spectral analysis; crosscorrelation analysis with the watermarking noise, etc.

Regarding unconditional security, that is, in the case of unlimited Trudy, the situation is different, see below.

**4. Cracking the watermarked security**

It is easy to realize that the watermarking security outlined above is not unconditional. Below we will focus on the linear response situation which is also assumed in the original papers [4,5] introducing watermarking to smart grids. It is the simplest and most practical case, however, the problem of nonlinear response is also interesting from a fundamental point of view, see Section 4.2.

While the sensors are tamper resistant (due to proper design or alarm system), the communication lines are not authenticated unconditionally. We illustrate the attack by Trudy targeting the line of Sensor-1; see Figures 3 and 4.

*4.1 Cracking the watermarked grid in the linear response case*

The signal $R_{1\text{DT}}(t)$ of Sensor-1 simulated by the Digital Twin can be written as:

$$R_{1\text{DT}}(t) = R_1(t) = S_1(t) - N_1(t) \ . \tag{1}$$

Thus the watermarking component $N_1(t)$ at Sensor-1 can be determined by the difference of the real and the simulated sensor signal:

$$N_1(t) = S_1(t) - S_{1DT}(t) \ . \tag{2}$$

Now, as the watermarking component is separated, Trudy can easily synthesize a new, faked and properly watermarked sensor signal:

$$S_{1f}(t) = R_{1f}(t) + KN_1(t) = R_{1f}(t) + K\left[S_1(t) - S_{1DT}(t)\right] \ , \tag{3}$$

where $K$ is the required scaling factor of the watermarking signal component. The value of $K$ must be determined case-by-case.

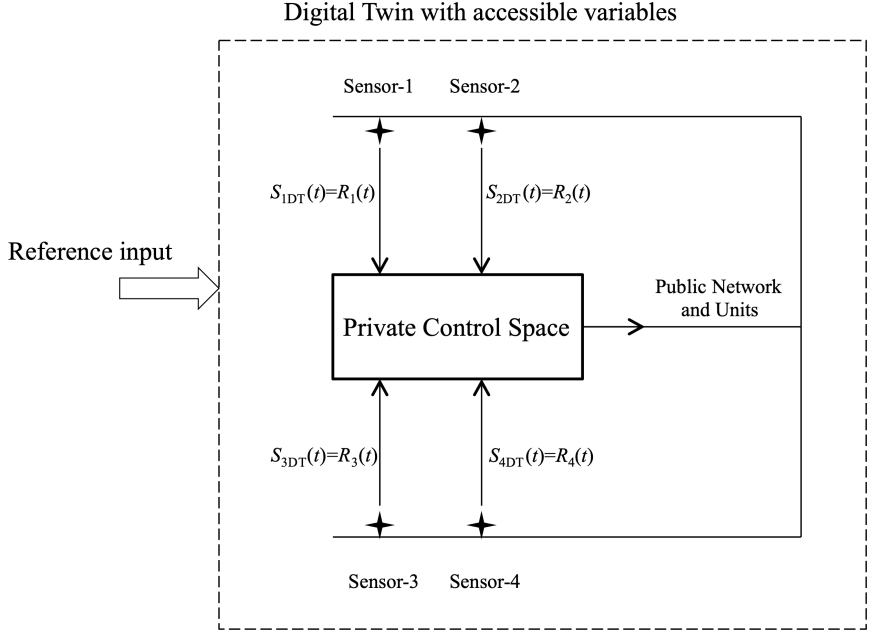

Figure 3. Illustration of the Digital Twin used for an attack. The simulated signal $S_{1DT}(t)$ of Sensor-1 is identical with the sensor signal $R_1(t)$ without watermarking. This fact is utilized by Trudy to extract the watermarking component $N_1(t)$ in the signal of Sensor-1, and to synthesize a fake watermarked signal see, an example in Figure 4.

Two simple examples for the proper choices of $K$:

(a) If the $R_1(t)$ sensor signal is independent from the voltage on the line, such as, when the sensor is a thermometer for identifying failing transformers, then

$$K=1, \tag{4}$$

because the watermarking signal $N$ remains to have the same RMS value even if and independent transformer heats up.

(b) If the $R_1(t)$ sensor signal's absolute amplitude scales linearly with the RMS value of the watermarking signal, for example, when the $R_1(t)$ represents the RMS voltage on the power line, the watermarking signal must scale with the line voltage, that is:

$$K = \frac{R_{1f}(t)}{R_1(t)} > 0 \ , \tag{5}$$

where the time coordinate *t* stands for the variations of the RMS voltage.

For an example of a malicious attack on the first sensor in the simplest situation with $K=1$, see Figure 4, where the original signal $R_1(t)$ is replaced by the fake one $R_{1f}(t)$, while the watermarking component $N_1(t)$ does not indicate any attack.

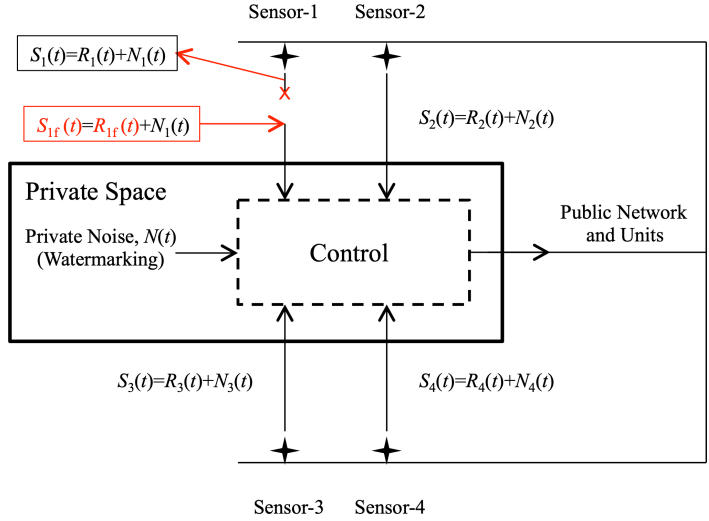

Figure 4. Example for cracking in the simplest situation, where $K=1$, see Equation 3.

Additional notes:

a) Alternatively, Trudy may try utilizing the inverse of the linear response to extract the watermarking noise and utilize that for creating a fake response. However, when there is significant time delay in the system, it may make the realization of such protocol difficult. This topic requires further study involving practical grid characteristics.

b) If the grid has inherent background noises, that is not a problem because the total noise can be treated in the same way as the watermarking noise discussed above. The described attack requires the extraction and usage of the total noise only. No separation of the background noise and the watermarking noise is needed. The protocol remains the same in such a case.

*4.2 Unconditionally authenticated communications with the sensors*

It is important to note that, if there is unconditionally secure key exchange between the sensors and the controller then unconditionally authenticated communications can be used for transferring the sensor data. Then the attacks described above do not work. However, then the need for watermarking is questionable.

*4.3 On the nonlinear response situation*

The works introduced watermarking for the smart grid [4-6] assumed linear response. Yet, the nonlinear response case is interesting, at least academically. Even though the smart grid control response is typically very near to linear, in special cases, Alice/Bob may attempt to utilize the nonlinear components of the response to enhance security. To avoid that, Trudy can use a more calculation-intensive protocol, which is similar to the perturbation calculus in quantum mechanics by utilizing the situation requested by practical grid operations, that:

$$\left\langle X_i^2(t) \right\rangle \ll \left\langle N_i^2(t) \right\rangle \ll \left\langle R_i^2(t) \right\rangle , \qquad (6)$$

which describes the near-to-linear situation with small watermarking signal.

Alternatively, Trudy can prepare in advance a lookup table between various values and amplitude paths between $N(t)$, $N_i(t)$, $R_i(t)$, and $X_i(t)$. Then she can use the linear approach as first-order result to estimate $N(t)$. Then she can correct the fake sensor signal accordingly. Note, this cracking scheme may have difficulties with extreme situations such as strong nonlinearity and large response delays. This topic, too, requires further study involving practical grid characteristics, though the importance of this issue is only academic.

## 5. Conclusion

We have generalized the notion of unconditional security to smart grids and shown that watermarking cannot secure the smart grid unconditionally. An open question is if unconditionally authenticated (public) communications with tamper resistance of the sensors are enough to guarantee unconditional smart grid security, or the whole communication scheme must be unconditionally secure to exclude Trudy from seeing the data.

*arxiv*

9